\documentclass[preprint,showpacs,preprintnumbers,amsmath,amssymb]{revtex4}
\usepackage{graphicx}
\begin{document}




\title{Adatom Doping-Enriched Geometric and Electronic Properties of Pristine Graphene: a Method to Modify the Band Gap
\footnote{``This paper is dedicated to Professor Lou Massa on the occasion of his 
Festschrift: A Path through Quantum Crystallography". 
}}
\author{ Ngoc Thanh Thuy Tran$^1$, Dipendra Dahal$^2$, Godfrey Gumbs$^{2,3}$ 
and  Ming-Fa Lin$^1$ }
\affiliation{$^1$Department of Physics, National Cheng Kung University, Tainan 701, Taiwan\\
$^{2}$Department of Physics and Astronomy, Hunter College of the
City University of New York, 695 Park Avenue, New York, NY 10065, USA\\
$^{3}$Donostia International Physics Center (DIPC),
P de Manuel Lardizabal, 4, 20018 San Sebastian, Basque Country, Spain}

\date{\today}

\begin{abstract}
We have investigated the way in which the concentration and distribution of  adatoms affect
the geometric and electronic properties of graphene. Our calculations were based on the use of first principle under the density functional theory which reveal various types of $\pi$-bonding. The energy band structure
of this doped graphene material may be explored experimentally by employing angle-resolved 
photo-emission spectroscopy (ARPES) for electronic band structure measurements and scanning tunneling spectroscopy
(STS) for the density-of-states (DOS) both of which have been calculated and reported in this paper.  Our
calculations show that such adatom doping is responsible for the destruction or appearance  of the Dirac cone 
structure. 

 \end{abstract}

\vskip 0.2in

\medskip
\par
\maketitle

\section{Introduction}
\label{sec1}

\medskip
\par
  
Ever since the successful exfoliation of graphene was achieved, there has been
considerable effort  both theoretically \cite{leni1,leni2,leni3,leni4, Ke-Yan,KAndre} and experimentally \cite{KAndre,leni5,leni6,leni7}     
to exploit its electronic, photonic and mechanical properties. Because of its 
flexibility, robust strength, heightened intrinsic mobility, high thermal 
and electrical conductivity and zero band gap at the $K$ point, linear 
dispersion and massless fermions, scientists have envisioned   
potential application for graphene-based electronic devices in the 
semiconductor industry. Creating pristine graphene in commercial quantities 
with a specific band gap is a major challenge  in its use for nanoelectronic 
devices.   Some of the intriguing properties which have received a  great 
deal of attention include the doping and temperature dependence of plasmon 
excitations and their energy dispersion dependence on the direction of the 
wave vector in the Brillouin zone  for graphene \cite{Wunsch,GG1,GG2,GG3,SDas,ds},  
the Klein paradox \cite{Beenakker, Allain, AHC, DD},  the Veselago lens 
\cite{Altshuler, DD, Vsl}, screened impurity potential \cite{GG4, Katsnelson}, 
effect of magnetic field \cite{GG4} to name just a few.

\medskip
\par

Placing a graphene layer in close proximity with a substrate such as silicon carbide, copper, hexagonal boron nitride 
or doping the graphene sheet with a variety of elements such as Hydrogen, Fluorine or Chlorine are some ways 
for designing the energy band structure of graphene in a specified  way.  Of these mentioned ways, this paper 
is primarily concerned  with an investigation of the properties of oxygenated graphene, also known as 
graphene oxide (GO). These GOs,  doped by oxygen atom either on one side or both sides,  are considered 
to be useful for a wide variety of device-based applications including supercapacitors \cite{leni8,leni9,leni10,leni11}, 
energy storage \cite{leni15,leni16},  sensors \cite{leni17,leni18,leni19}, mermistors   \cite{leni15}
 and water purification \cite{leni38}.

\medskip
\par

Manufacturing GO commercially at a reasonable cost and in an environmentally favorable manner
is another challenge for this material.   There exist several methods for producing GO.  In 
1859, Brodie \cite{leni39}  treated graphite with $KClO_3$, an oxidant, to oxidize graphite in 
$HNO_3$. Staudenmaier  \cite{leni40} improved this method later by   adding $H_2SO_4$ to increase the 
acidity of the mixture. Hummer  \cite{leni41} further developed this method by adding a mixture 
of $NaNO_3$, $H_2SO_4$ and $KMnO_4$. This latter method is the most widely used because of 
its shorter reaction time and no $ClO_3$ emission \cite{leni42,leni43}. Modification of 
Hummer's method is the addition of $H_3PO_4$ with $H_2SO_4$ in the absence of $NaNO_3$ 
and increasing the amount of $KMnO_4$ \cite{leni44}. The GOs have been manufactured using a
bottom-up method since this process is simpler and more acceptable from an 
environmental point of view compared with traditionally top-down methods. Different 
oxygen concentration may be synthesized by controlling the amount of oxidant $KMnO_4$. 
Also, GO has been manufactured using a bottom-up Tang-lau method since it is simpler and 
meets environmental standards compared with traditionally top-down methods \cite{leni45}. 
The concentration of oxygen may be controlled by regulating the amount of the oxidant 
compound $KMnO_4$\cite{leni46}, or the oxidation time.
	
\medskip
\par

Our aim in this paper is to shed light on the electronic properties and 
geometrical structure of graphene oxide with the use of the Vienna Ab initio Simulation 
Package (VASP) \cite{kresse1996efficient}. The exchange-correlation energy due to the electron-electron interactions is calculated from the Perdew-Burke-Ernzerhof functional under the generalized gradient approximation \cite{leniadded1} . The projector-augmented wave pseudopotentials are employed to evaluate the electron-ion interactions \cite{leniadded2}.  A plane-wave basis set with a maximum plane-wave energy of 500 eV is used for the valence electron wave functions. The convergence criterion for one full relaxation is mainly determined by setting the Hellmann-Feynman forces smaller than $0.01 eV/^\circ$A and the total energy difference less than $10^{-5} eV$. The first Brillouin zone is sampled in a Gamma scheme by $30\times30\times1$ k points for structure relaxations, and then by $100\times100\times1$ k points for further calculations on electronic properties. We present  our results for different concentrations of C/O. Also, we reported the DOS and energy band structure of GOs to address 
the critical issue of how to modulate the band gap of graphene. 
The outline of the rest of our presentation is as follows: In Sec.\ \ref{sec2},
 we present and discuss our results for the band structure of GO for various
adatom dopings, its geometrical structure and DOS. Section\ \ref{sec3}
is devoted to some concluding remarks.

\section{Results and Discussions}
\label{sec2}

\medskip
\par

\begin{figure}
\graphicspath{{figure}}
\centering
  \includegraphics[scale=2.7]{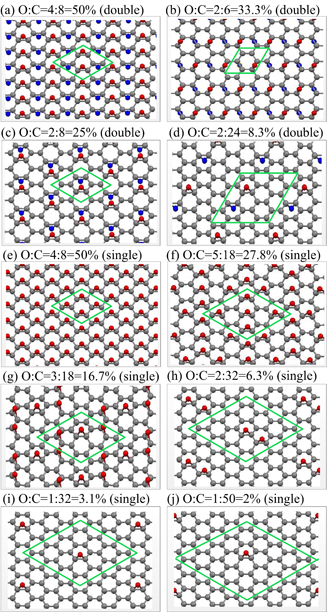}
\caption{(Color online) Geometric structures of GOs with various 
concentrations: double-side adsorptions (a) 50\%, (b) 33.3\%, (c) 25\%, 
and (d) 8.3\%; single-side adsorptions (e) 50\%, (f) 27.8\%, (g) 16.7\%, 
(h) 6.3\%, (i) 3.1\%, and (j) 2\%. The red and blue atoms, respectively, 
correspond to O atoms adsorbed on the top and bottom of a graphene layer (gray color).}
\label{FIG:1}
\end{figure}

The stable adsorption position of O atoms on graphene is determined by 
comparing the total ground state energies, in which the lower value 
corresponds to higher stability. Among the bridge-, hollow- and top-sites, 
our calculations show that the bridge is the most stable one and the top-site
is the least stable, in agreement with previous studies \cite{leni1}. 
After self-consistent calculations, we found that all the GO systems, with various 
concentrations and distributions, have stable structures with very slight buckling present
as demonstrated in Fig.\  \ref{FIG:1}. The main features of geometric structures, 
including the C-C and C-O bond lengths are dominated by the O-concentrations, as shown 
in Table 1. As oxygen atoms are adsorbed on graphene, the C-C bond lengths are 
dramatically expanded to 1.53 $\mbox\AA$ for the case of single-side adsorption 
($50$\%), which is close to $1.54$ $\mbox\AA$ sp$^3$ bond length in diamond. The 
C-C bond lengths gradually recover to that in pristine graphene (1.42 $\mbox\AA$) with 
the decrease of O-concentration, which is consistent with experimental \cite{KAndre} 
and theoretical \cite{discuss3}  results. However, the C-O bond lengths exhibit the 
opposite dependence. These results indicate that the stronger orbital hybridizations in C-O bonds significantly reduce their lengths and even weaken the $\sigma$ bonds in C-C bonds. There exists a partial transformation from sp$^2$ bonds in pristine graphene into sp$^3$ bonds, in agreement with previous studies \cite{KAndre, Hirata}. The above-mentioned rich geometric structures lead to diverse electronic properties.

\medskip
\par

\begin{figure}[h]
\graphicspath{{figure}}
\centering
 \includegraphics[scale=.6]{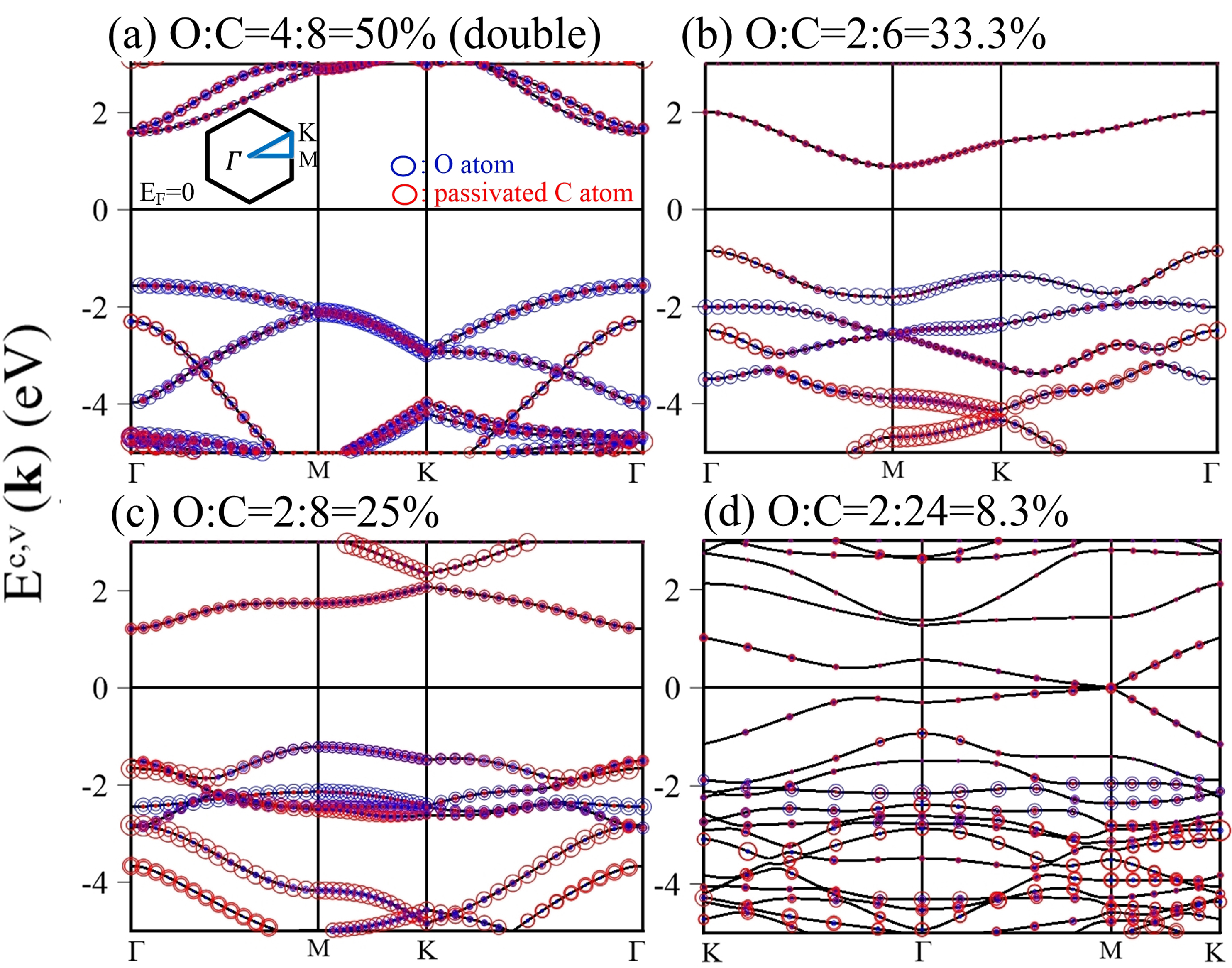}
\caption{(Color online)  Band structures of double-side adsorbed GOs with various concentrations: 
(a) 50\%, (b) 33.3\%, (c) 25\%, and (d) 8.3\%. Superscripts c and v correspond to the conduction 
and valence bands, respectively. The red and blue circles, respectively, correspond to the contributions 
of passivated C atoms and O atoms, in which the dominance is proportional to the radius of circles. Also 
shown in the inset of (a) is the first Brillouin zone.}
\label{FIG:2}
\end{figure}

\begin{figure}[h]
\graphicspath{{figure}}
\centering
 \includegraphics[scale=.6]{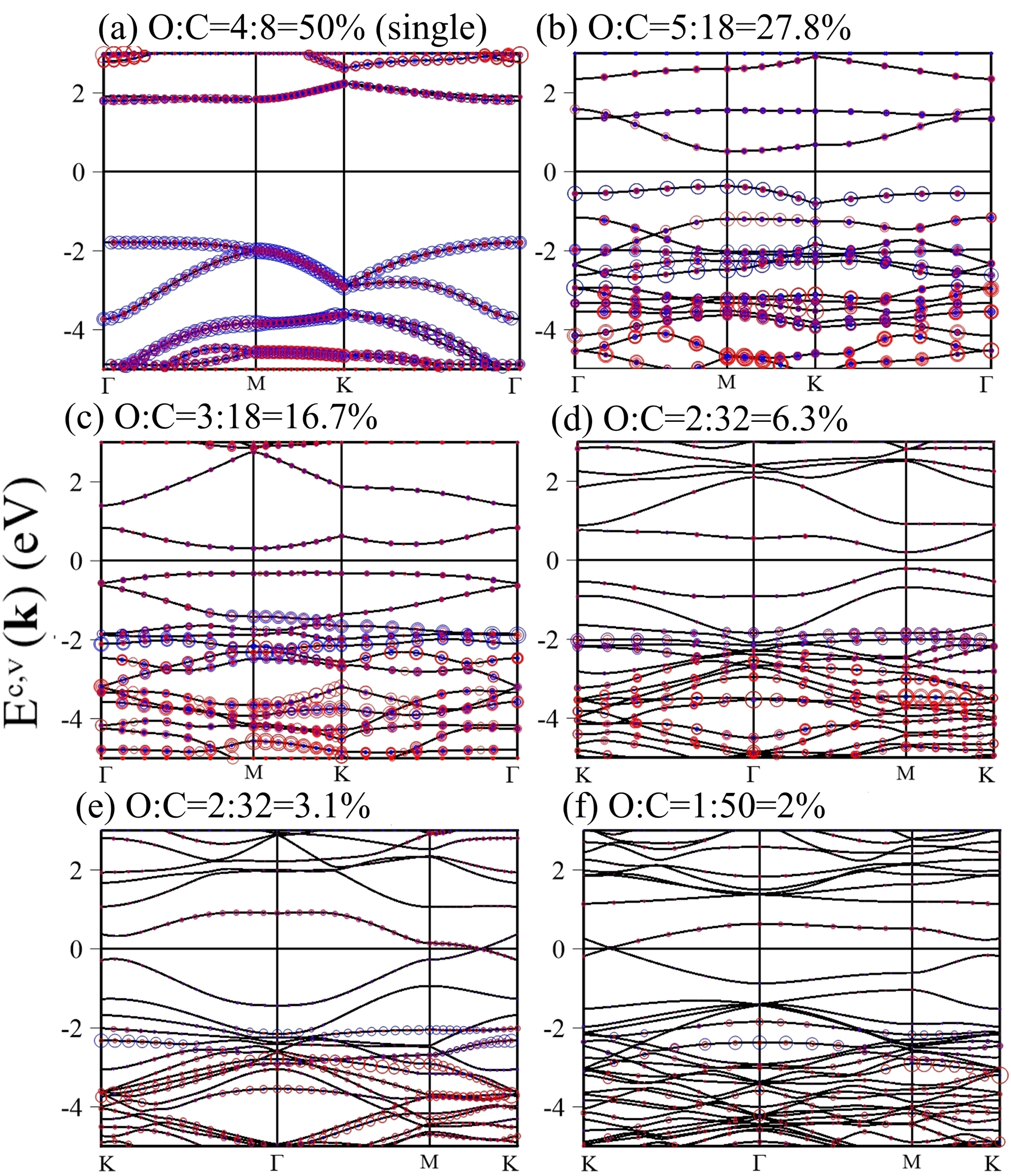}
\caption{(Color online)  Band structures of single-side adsorbed GOs with various concentrations: 
(e) 50\%, (f) 27.8\%, (g) 16.7\%, (h) 6.3\%, (i) 3.1\%, and (j) 2\%. }
\label{FIG:3}
\end{figure}

The two-dimensional (2D) band structures along with high symmetry points 
are useful for examining the electronic properties. When oxygen atoms are 
adsorbed on graphene, the energy dispersion relations exhibits dramatic
changes with the variations in O-concentrations and O-distributions. For 
monolayer graphene, the sp$^2$ orbitals of (2p$_\text{x}$, 2p$_\text{y}$, 2s) 
possess very strong covalent $\sigma$-bonds among the three nearest-neighbor 
carbon atoms, while the 2p$_\text{z}$ orbitals contribute to the weak 
$\pi$-bonds, being associated with the Dirac-cone structure right at the 
Fermi-level ($E_F$). The $\sigma $ and $\sigma^\ast$ bands form the 
high-lying electronic structures within a range of $|E^{c ,v}|\ge$ 3 eV, 
in which the extremum energies are at the $K$ point. The band structures 
of double-side and single-side adsorptions GOs are shown in Fig. 2 and 
Fig.\ \ref{FIG:3}, respectively. The contributions of O atoms and C atoms 
passivated with the former are represented by blue and red circles, respectively, 
in which the dominance is proportional to the circle's radius. Different 
from pristine graphene, the isotropic Dirac-cone structure near the $K$ point 
is destroyed, mainly owing to the serious hybridization between the atomic 
orbitals of O and passivated C atoms (C atoms bonded O). Instead, there 
are wide energy gaps (Figs. \ref{FIG:2}(a)-(c) and Fig. \ref{FIG:3}(a)) 
and O-dominated energy bands nearest to $E_F$ (blue circles). That is, 
the $\pi$ bonds due to the neighboring bonds of parallel 2$p_z$ orbitals 
are replaced by the orbital hybridizations of O-O bonds. As to the strong 
orbital hybridizations of passivated C atoms and O atoms (red and blue circles 
dominate simultaneously), their energy bands are located in 
-2.5 eV$\,\le\,E^v\le\,$-4 eV. The deeper $\sigma$-bands with $E^v\le\,$-4 eV 
are formed by the rather strong hybridization of (2$p_x$, 2$p_y$, 2s) 
orbitals in C-C bonds almost independent of O-adsorption. With the 
decrease of O-concentrations, the O-dominated bands become narrower 
and contribute at deeper energy, while the $\pi$ bands are gradually 
recovered (Fig.\ \ref{FIG:2}(d) and Figs. \ref{FIG:3}(b)-(f)). This 
leads to a reduced band gap and the reformation of a distorted Dirac-cone structure.

\medskip
\par

The energy gap ($E_g$) values are sensitive to changes in the O-concentration 
and -distribution, as listed in Table 1. They could be divided into three 
categories, although the dependence on the O-concentration is non-monotonic
\cite{leni2}. The GO systems always have a finite gap for O-concentrations 
higher than 25\%, mainly owing to the sufficiently strong O-O and C-O bonds. 
In addition, $E_g$ values are also affected by double-side or single-side 
adsorptions. Compared to the double-side adsorption of 50\% O-concentration, 
the single-side ones possess larger band gaps, mainly owing to the stronger 
interactions between O atoms. In the concentration range of 25-4\%, the 
$E_g$ values decline quickly. This might exhibit small or vanishing gaps 
related to the reformed distorted Dirac cones Fig. \ref{FIG:2}(d) and 
Figs. \ref{FIG:3}(c)-(d)). The energy gap becomes zero for low concentrations 
of $<$4\%. This is induced by the fully reformed Dirac cones without energy 
spacing (Figs. \ref{FIG:3}(e)-(f)). These feature-rich energy bands of GOs, 
including the absence and presence of the distorted Dirac-cone structures, 
and the band gap are expected to be examined by ARPES. By controlling the O-coverage, the band gap of 
graphene may be significantly adjusted, which opens up the scope for 
potential use in nanoelectronic devices.

\medskip
\par

\begin{figure}[h]
\graphicspath{{figure}}
\centering
 \includegraphics[scale=.6]{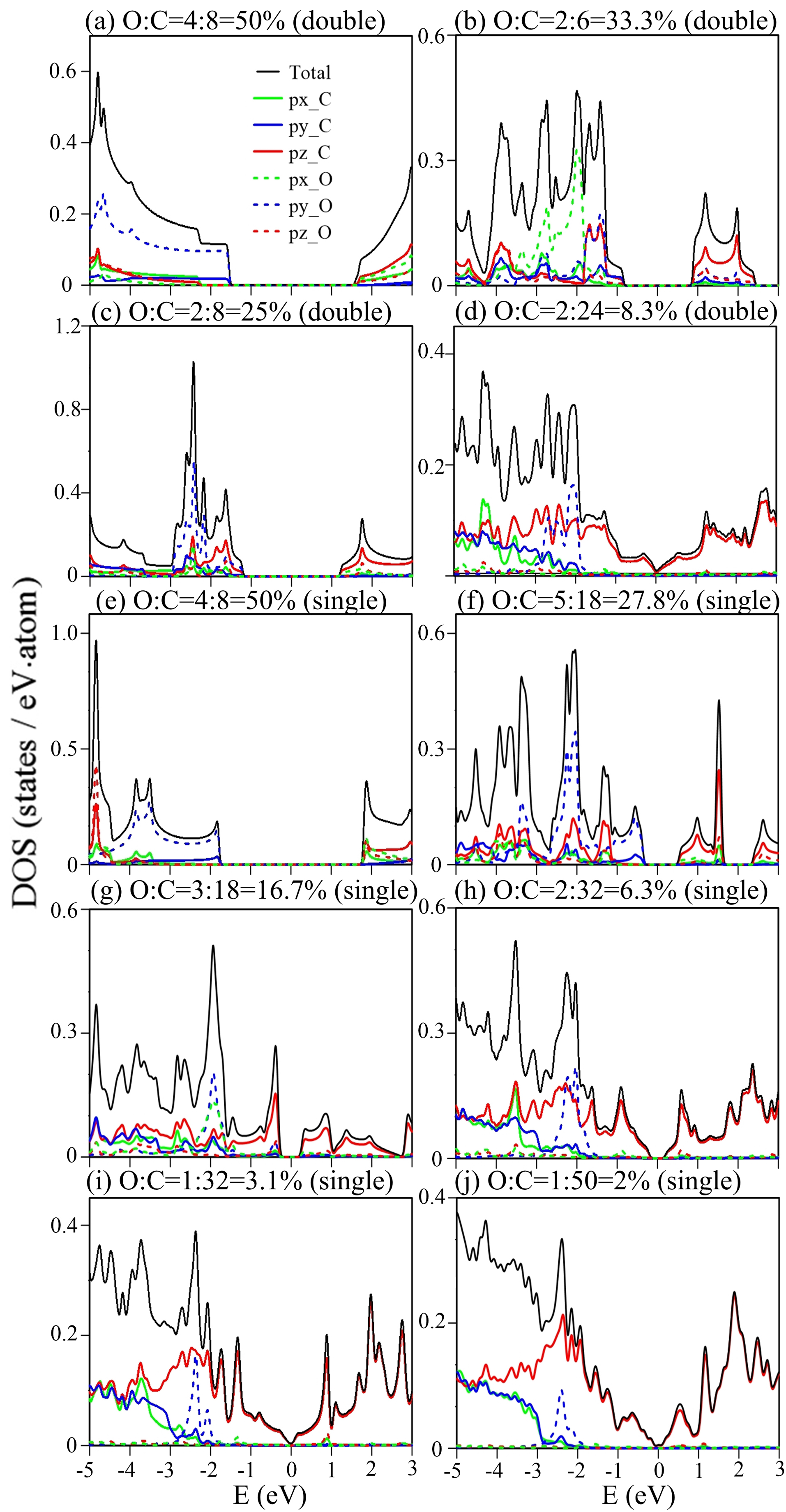}
\caption{ (Color online)  Orbital-projected DOSs of GOs with various O-concentrations:(a) 50\%, (b) 33.3\%, 
(c) 25\%, (d) 8.3\%, (e) 50\%, (f) 27.8\%, (g) 16.7\%, (h) 6.3\%, (i) 3.1\%, and (j) 2\%. Figs. (a)-(d) belong 
to double-side adsorptions, whereas Figs. (e)-(j) belong to single-side ones. }
\label{FIG:4}
\end{figure}

The main characteristics of the band structures are directly reflected 
in the DOSs, as shown in Fig.\ \ref{FIG:4}. The predicted C-O and O-O 
bonds as a results of oxygen adsorption are clearly viewed by the 
orbital-projected DOSs. The low-energy DOS is dramatically altered after 
oxygens are adsorbed on graphene. For pristine graphene, the peaks caused 
by $\pi$ and $\pi^\ast$ bands due to 2$p_z$-2$p_z$ bond between C atoms 
will dominate within the range of -2 eV$\,\le\,|E|\le\,$2 eV. Moreover, a 
vanishing gapless DOS near $E_F=$0 indicates the zero-gap semiconducting 
behavior. However, for high O-concentrations, the $\pi$ and $\pi^*$-peaks 
are absent as a result of the strong C-O bond. Instead, there are energy gaps
and several O-dominated prominent structures in a wide range of $E^v\le\,$-2.5 
to $E_F$ (Figs.\ \ref{FIG:4}.(a)-(c); Figs. \ref{FIG:4}(e)-(f)). Also, 
the magnitude of the energy gap depends on the competition between the 
O-O bonds and the $\pi$ bonds. Concerning the middle-energy range of 
-2.5 eV$\,\le\,E^v\le\,$-4 eV, the DOSs grow quickly and exhibit prominent 
peaks due to the C-O bonds. With the decrease in the O-concentration, 
the $\pi$ and $\pi^*$ -peaks are gradually reformed at low energy, while 
the O-dominated structures present a narrower width and large red shift 
deeper from the fermi level (Fig.\ \ref{FIG:4}(d); (Figs. \ 
\ref{FIG:4}(g)-(j)). The critical features in DOSs, mainly the finite 
or vanished energy gaps, the $\pi$- and $\pi^\ast$-peaks, the O-dominated 
special structures, and the (C,O)-dominated prominent peaks can be further 
verified by STS, giving useful information 
in determining the oxygen concentration.

\medskip
\par

\section{Concluding Remarks}
\label{sec3}

\medskip
\par

 In summary, an important result coming out of our study is that the 
eletronic and geometric structure of graphene may be modified by 
oxygenation either from one side or from both side. Depending on the 
distribution and the concentration of the adsorbed oxygen atom, 
the bond lengths between C-C and C-O change. This leads to the reconstruction 
of the electronic structure causing the energy band structure to be modified. 
A finite energy gap emerges as a result of doping and the value of this 
band gap increases if the level of doping is increased. The emerged 
energy band gap also depends on single side or double side oxygen 
adsorption. This behaviour in fact plays an important role in tuning 
the band gap and modulating the energy band structure of GOs which 
could be used in designing the on-off feature in semiconductor devices.

\newpage

\begin{table}[htb]
\small
\caption{The optimized C-O bond lengths, C-C bond lengths and energy 
gap ($E_g$) for various concentrations of GOs.}
\label{t1}
\begin{tabular}{ c c c c c c}
Adsorption & O:C & \% & C-O bond & C-C bond & $E_g$ \\
& & & length (\AA)& length (\AA)& (eV)\\
\hline
Double-side & 4:8 & 50 & 1.44 & 1.49 & 3.16 \\
& 2:6 & 33.3 & 1.47 & 1.45 & 1.72\\
& 2:8 & 25 & 1.48 & 1.47 & 2.44\\
& 2:24 & 8.3 & 1.49 & 1.45 & 0\\
Single-side & 4:8 & 50 & 1.43 & 1.53 & 3.56\\
& 5:18 & 27.8 & 1.46 & 1.47 & 0.88\\
& 3:18 & 16.7 & 1.48 & 1.45 & 0.64 \\
& 2:32 & 6.3 & 1.48 & 1.43 & 0.36\\
& 1:32 & 3.1 & 1.48 & 1.42 & 0\\
& 1:50 & 2 & 1.48 & 1.42 & 0\\
\hline
\end{tabular}
\end{table}

\newpage

\end{document}